\documentclass[preprint2]{aastex}
\usepackage{graphicx}
\usepackage{epsfig}
\shorttitle{Physical Structure of NGC\,3242}
\shortauthors{Ruiz et al.}

\begin{document}

%% LaTeX will automatically break titles if they run longer than
%% one line. However, you may use \\ to force a line break if
%% you desire.

\title{
Physical Structure of the Planetary Nebula NGC 3242 from the Hot Bubble 
to the Nebular Envelope\footnote{
Based on observations obtained with \emph{XMM-Newton}, an 
ESA science mission with instruments and contributions directly 
funded by ESA Member States and NASA.}
}

%% Use \author, \affil, and the \and command to format
%% author and affiliation information.
%% Note that \email has replaced the old \authoremail command
%% from AASTeX v4.0. You can use \email to mark an email address
%% anywhere in the paper, not just in the front matter.
%% As in the title, use \\ to force line breaks.

\author{Nieves Ruiz\altaffilmark{1}, Mart\'{\i}n A.\ Guerrero\altaffilmark{1}}
\affil{$^1$ Instituto de Astrof\'{\i}sica de Andaluc\'{\i}a, CSIC, 
            Granada 18008, Spain}
\email{nieves@iaa.es, mar@iaa.es}

\and

\author{You-Hua Chu\altaffilmark{2}, Robert A. Gruendl\altaffilmark{2}}
\affil{$^2$ Astronomy Department, University of Illinois at 
            Urbana-Champaign, Urbana, IL 61801}
\email{yhchu@astro.illinois.edu, gruendl@astro.illinois.edu }

%% Notice that each of these authors has alternate affiliations, which
%% are identified by the \altaffilmark after each name.  Specify alternate
%% affiliation information with \altaffiltext, with one command per each
%% affiliation.

%%\altaffiltext{1}{Visiting Astronomer, Cerro Tololo Inter-American Observatory.
%%CTIO is operated by AURA, Inc.\ under contract to the National Science
%%Foundation.}
%%\altaffiltext{2}{Society of Fellows, Harvard University.}
%%\altaffiltext{3}{present address: Center for Astrophysics,
%%    60 Garden Street, Cambridge, MA 02138}
%%\altaffiltext{4}{Visiting Programmer, Space Telescope Science Institute}
%%\altaffiltext{5}{Patron, Alonso's Bar and Grill}

%% Mark off your abstract in the ``abstract'' environment. In the manuscript
%% style, abstract will output a Received/Accepted line after the
%% title and affiliation information. No date will appear since the author
%% does not have this information. The dates will be filled in by the
%% editorial office after submission.

\begin{abstract}

One key feature of the interacting stellar winds model of the formation 
of planetary nebulae (PNe) is the presence of shock-heated stellar wind 
confined in the central cavities of PNe.  
This so-called hot bubble should be detectable in X-rays.  
Here we present \emph{XMM-Newton} observations of NGC\,3242, a 
multiple-shell PN whose shell morphology is consistent with
the interacting stellar winds model.
Diffuse X-ray emission is detected within its inner shell with 
a plasma temperature $\sim$2.35$\times$10$^6$~K and an intrinsic 
X-ray luminosity $\sim$2$\times$10$^{30}$ ergs~s$^{-1}$ at the 
adopted distance of 0.55 kpc.  
The observed X-ray temperature and luminosity are in agreement 
with ``ad-hoc'' predictions of models including heat conduction.  
However, the chemical abundances of the X-ray-emitting plasma seem to imply 
little evaporation of cold material into the hot bubble, whereas the thermal 
pressure of the hot gas is unlikely to drive the nebular expansion as it is 
lower than that of the inner shell rim.  
These inconsistencies are compounded by the apparent large filling 
factor of the hot gas within the central cavity of NGC\,3242.

\end{abstract}

%% Keywords should appear after the \end{abstract} command. The uncommented
%% example has been keyed in ApJ style. See the instructions to authors
%% for the journal to which you are submitting your paper to determine
%% what keyword punctuation is appropriate.


%% From the front matter, we move on to the body of the paper.
%% In the first two sections, notice the use of the natbib \citep
%% and \citet commands to identify citations.  The citations are
%% tied to the reference list via symbolic KEYs. The KEY corresponds
%% to the KEY in the \bibitem in the reference list below. We have
%% chosen the first three characters of the first author's name plus
%% the last two numeral of the year of publication as our KEY for
%% each reference.

%% Authors who wish to have the most important objects in their paper
%% linked in the electronic edition to a data center may do so by tagging
%% their objects with \objectname{} or \object{}.  Each macro takes the
%% object name as its required argument. The optional, square-bracket 
%% argument should be used in cases where the data center identification
%% differs from what is to be printed in the paper.  The text appearing 
%% in curly braces is what will appear in print in the published paper. 
%% If the object name is recognized by the data centers, it will be linked
%% in the electronic edition to the object data available at the data centers  
%%
%% Note that for sources with brackets in their names, e.g. [WEG2004] 14h-090,
%% the brackets must be escaped with backslashes when used in the first
%% square-bracket argument, for instance, \object[\[WEG2004\] 14h-090]{90}).
%%  Otherwise, LaTeX will issue an error. 

\section{Introduction}

Planetary Nebulae (PNe) consist of stellar material ejected by low- 
and intermediate-mass stars (0.8--1.0~M$_\odot\le M_i \le$\,8--10~M$_\odot$).  
Towards the end of the Asymptotic Giant Branch (AGB), 
these stars experience copious mass loss and eject 
most of their stellar envelope through a slow, dense AGB wind.
The ejected material is subsequently ionized by the central star 
and becomes a PN. 
PNe eventually disperse into the interstellar medium as they 
expand, whereas the stellar cores, mainly composed of carbon 
and oxygen, will evolve toward the white dwarf stage.

Near the time when the hot stellar core is exposed, the slow AGB wind, 
with terminal velocities 5--30 km s$^{-1}$ \citep{ELT88}, is superseded 
by a fast stellar wind with terminal velocities 1,000--4,000 km s$^{-1}$ 
\citep{CRP85,GR-LM10}.  
This fast stellar wind sweeps up the slower AGB wind to form a PN 
\citep{Kwok83}.
In this interacting stellar winds (ISW) model, the physical structure of 
a PN would be similar to that of a wind-blown bubble, as modeled by 
\citet{WR77}, comprising a central cavity filled with shocked fast 
wind (the so-called hot bubble), a dense shell of swept-up AGB wind at 
10$^{4}$ K, and an outer envelope of unperturbed expanding AGB wind.  
In a simplistic model, the temperature of the shocked stellar wind 
inside the hot bubble would be 10$^{7}$-10$^{8}$ K, but turbulent 
mixing \citep[e.g.,][]{MF95} or heat conduction \citep{ZP98,SSW08} 
lowers the temperature of the hot gas to 10$^{6}$-10$^{7}$ K and raises 
its density to produce optimal conditions for the emission of soft 
X-rays. 
% In this interacting stellar winds (ISW) model, the physical structure of 
% a PN is similar to that of a wind-blown bubble, as modeled by \citet{WR77}. 
% The PN comprises a central cavity filled with shocked fast wind at 
% temperatures of 10$^{7}$-10$^{8}$ K (the so-called hot bubble), a 
% dense shell of swept-up AGB wind at 10$^{4}$ K, and an outer 
% envelope of unperturbed expanding AGB wind \citep{MF95,ZP98,SSW08}. 
Therefore, X-ray observations of shock-heated hot gas in PNe provide 
us a direct means to examine the interaction of the fast stellar wind 
with the AGB wind and to investigate the transfer of energy and momentum 
to the PN envelope.

\emph{ROSAT} observations showed hints of diffuse X-ray emission 
in a few PNe \citep{GR00}. 
However, it was not until the advent of \emph{Chandra} and 
\emph{XMM-Newton}, with their unprecedented resolution and sensitivity, 
that we were finally able to unambiguously detect hot gas in PNe. 
\emph{Chandra} and \emph{XMM-Newton} have resolved the diffuse X-ray emission 
in a handful of PNe \citep[e.g.,][]{KN00,KN01,CH01,GR02,GR05} and revealed 
unexpected, hard X-ray emission from the central stars of several PNe that may 
originate from the coronal emission of unseen faint binary companions or shocks
within the fast stellar winds \citep{GR01,KN03,Montez_etal2010}.  
Observations of diffuse X-ray emission in PNe demonstrate that hot 
gas in elliptical PNe is confined within the innermost nebular shell 
and that the high pressure of the hot gas may indeed drive the nebular 
expansion as expected in bubble models.

The observed values of $L_{\rm X}$ and $T_{\rm X}$ are in agreement with 
predictions of the time-dependent models developed by \citet{SSW08} that
include heat-conduction for PNe with central stars of normal, hydrogen-rich 
surface composition.
There is, however, an unsolved puzzle: the analyses of the chemical 
composition of the X-ray-emitting plasma in PNe suggest that it is 
mainly composed of stellar wind, with little contamination of material 
from the cold nebular shell \citep[e.g., NGC\,2392 and 
NGC\,6543,][]{GR05,CH01}.  
This discrepancy has been further illustrated by the analysis of the 
high-spectral resolution \emph{Chandra} LETG spectrum of BD+30\degr3639 
that conclusively confirms the stellar wind composition of its 
X-ray-emitting plasma \citep{Yu_etal2009}.  
This problem has prompted alternative mechanisms for the production 
of hot gas in PNe, including the action of fast collimated outflows 
and/or slow fragments in the onset of the fast stellar wind 
\citep{SK03}, and the absorption of energy from the stellar wind by 
slowly moving ions embedded in the wind itself, the so-called pick-up 
ions \citep{Soker_etal2010}.

NGC\,3242 (PN\,G261.0+32.0), the Ghost of Jupiter, is a multiple-shell 
PN with a bright, 28\arcsec$\times$20\arcsec\ inner ellipsoidal shell and 
ansae surrounded by a fainter, 46\arcsec$\times$40\arcsec\ moderately 
elliptical envelope.  
These two shells are further enclosed by arcs and a giant 
broken halo revealed by deep images \citep{COR03,COR04}.
The double-shell morphology of the main nebula of NGC\,3242 is highly 
suggestive of interactions between the fast stellar wind of its central 
star \citep[$v_{\infty}$=2,400 km~s$^{-1}$,][]{PD04} and the previous 
slow AGB wind.  
The AGB wind (the nebular envelope) has been swept by the fast 
stellar wind to form a thin ionized shell with a central cavity 
that can be expected to be filled with shocked fast wind.
This shock-heated gas should emit X-rays, and the diffuse X-ray emission
from NGC\,3242 is likely detectable because of its proximity 
\citep[distance = 0.55$\pm$0.23 kpc,][]{TZ97,ML04}, and low extinction 
\citep{Balick_etal93,HE00,POT08}.

In this paper, we present \emph{XMM-Newton} observations of NGC\,3242 that 
have detected diffuse X-ray emission within its innermost nebular shell.  
The observations are described in \S2, the results are presented 
in \S3, the physical structure of the optical nebula is investigated 
in \S4, and the effects of the shocked stellar wind in the nebula 
are discussed in \S5.

\section{Observations}

\subsection{XMM-Newton X-Ray Observations}

NGC\,3242 was observed with the \emph{XMM-Newton} Observatory in 
Revolution 730 on 2003 December 4 using the EPIC-MOS1, EPIC-MOS2, 
and EPIC-pn CCD cameras (OBSID = 0200240401).
The two EPIC-MOS cameras were operated in the Full-Frame Mode for 
a total exposure time of 19.1 ks, while the EPIC-pn camera was 
operated in the Extended Full Frame Mode for a total exposure time 
of 15.7 ks. 
The Medium filter was used for all observations. 
The \emph{XMM-Newton} products were processed using the 
\emph{XMM-Newton Science Analysis Software} (SAS version 
10.0.0) and the calibration files from the Calibration 
Access Layer available on 2010 September 16. 
The event files were screened to eliminate events due to charged 
particles or associated with periods of high background. 
For the EPIC-MOS observations, only events with CCD patterns 0--12 
were selected; for the EPIC-pn observation, only events with CCD 
pattern 0 (single pixel events) were selected. 
Time intervals of high background, when the count rate in the background 
dominated 10--12 keV energy range is $\ge$0.3 cnts~s$^{-1}$ for EPIC-MOS 
and $\ge$1.4 cnts s$^{-1}$ for EPIC-pn, were discarded.
The resulting net exposure times are 18.7 ks, 18.7 ks, and 13.6 ks 
for the EPIC-MOS1, EPIC-MOS2, and EPIC-pn observations, respectively.

The \emph{XMM-Newton} EPIC observations detect a source of diffuse 
X-ray emission at the location of NGC\,3242.  
An inspection of EPIC-pn and EPIC-MOS images at different energy ranges 
reveals that this source is soft, with most emission below 1.0 keV and 
very little emission at higher energies. 
The EPIC-pn background-subtracted count rate in the 0.38-2.0 keV 
energy range is 31.3$\pm$1.6 cnts~ks$^{-1}$ for a total 
of 422$\pm$22 counts.
The EPIC-pn net count rates in the energy ranges 0.38-1.0 keV 
and 1.0-2.0 keV are 30.3$\pm$1.6 cnts~ks$^{-1}$, and 
1.1$\pm$0.4 cnts~ks$^{-1}$, respectively.  
The EPIC-MOS background-subtracted count rates in the 0.38-2.0 keV energy 
band are 4.4$\pm$0.5 cnts~ks$^{-1}$ for MOS1 and 4.8$\pm$0.5 cnts~ks$^{-1}$ 
for MOS2, with a total of 83$\pm$10 cnts for MOS1 and 90$\pm$10 cnts for 
MOS2.

\subsection{Archival Narrow-band HST Imaging }

In order to examine the spatial correlation between the diffuse X-ray 
emission and the optical nebula, narrow-band WFPC2 images of NGC\,3242 
in the H$\alpha$, He~{\sc ii} $\lambda$4686, [N~{\sc ii}] $\lambda$6583, 
and [O~{\sc iii}] $\lambda$5007 emission lines were retrieved from the 
\emph{HST} archive (Proposal ID 7501 and 8773, PI: Arsen Hajian, and 
Proposal ID 6117, PI: Bruce Balick).  
The images used in this work are listed in Table~\ref{tab.hst} with their 
integration times and filters. 
In all cases, the innermost shell of the nebula was registered 
on the WFPC2-PC1 CCD, while a fraction of the outer envelope 
was missed by the detector.  
These images were calibrated via the pipeline procedure and cosmic 
rays were removed by combining different exposures obtained with the same 
filter using standard IRAF\footnote{
IRAF, the Image Reduction and Analysis Facility, is distributed by 
the National Optical Astronomy Observatory, which is operated by 
the Association of Universities for Research in Astronomy (AURA) 
under cooperative agreement with the National Science Foundation.} 
routines. 
The final images have total exposure times of 100 s for H$\alpha$, 
160 s for He~{\sc ii}, 1260 s for [O~{\sc iii}], and 2840 s for 
[N~{\sc ii}].

\subsection{Medium-Dispersion Spectroscopy Observations}

Medium-dispersion long-slit spectroscopic observations of NGC\,3242, 
obtained using the GoldCam CCD Spectrograph (GCCAM) on the 2.1m 
telescope at the Kitt Peak National Observatory (KPNO) on 1996 
December 7, were kindly provided to us by Dr.\ K.\ Kwitter \citep{HE00}. 
The observations, covering the spectral region $3600-9600$ \AA, were 
obtained using a 285\arcsec$\times$5\arcsec\ slit oriented along the 
East-West direction and centered at 8\arcsec\ south of the central star.
The spatial scale of the observations is 0\farcs78~pix$^{-1}$. 
The grating 240 was used with the GG-345 blocking filter to cover the blue
spectral region $3650-6750$ \AA\ at a spectral dispersion 1.49 \AA~pix$^{-1}$, 
whereas the red spectral region $5650-9600$ \AA\ was observed at a spectral 
dispersion 1.86 \AA~pix$^{-1}$ using grating 58 with the OG-530 blocking 
filter.  
Two exposures of 60 s each were acquired in the blue spectral range, 
while exposures of 60 s, 30 s, and two of 20 s were acquired in the 
red spectral range. 
The original spectra were reduced and analyzed using standard IRAF 
routines.  
For the flux calibration we used several observations of the 
spectrophotometric standard stars BD+28$^\circ$4211 and Hiltner\,600 
obtained on the same night.

\section{XMM-Newton Data Analysis}

\subsection{Spatial Properties of the X-ray Emission from NGC\,3242}

\begin{figure*}[Ht!]
 \begin{center}
\includegraphics[width=16.5cm]{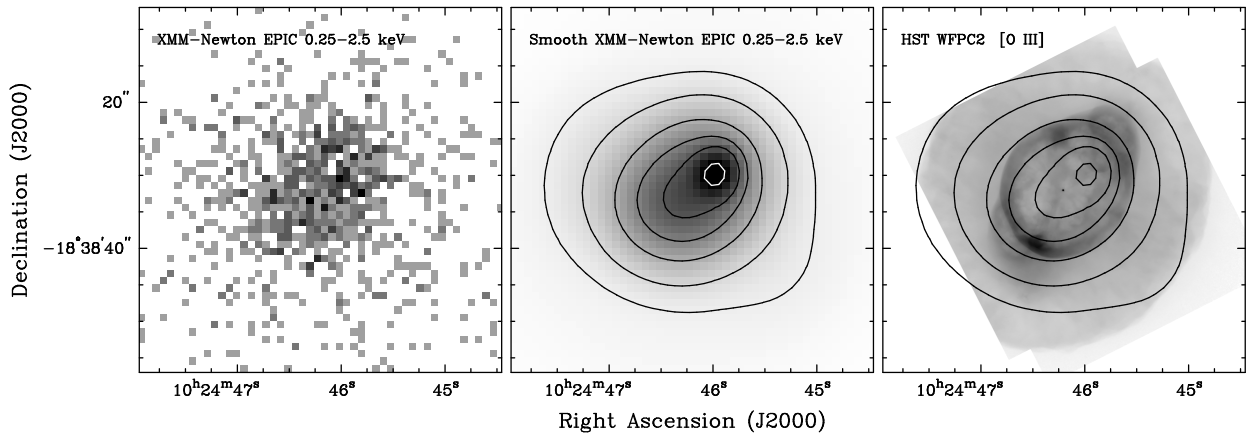}
 \end{center}
 \caption{
Raw ({\it left}) and smoothed ({\it center}) \emph{XMM-Newton} EPIC 
images of NGC\,3242 in the 0.25-2.5 keV energy band, and ({\it right}) 
\emph{HST} WFPC2-PC1 [O~{\sc iii}] image of NGC\,3242.  
The [O~{\sc iii}] image is overplotted with the X-ray contours 
derived from the smoothed EPIC X-ray image.  
Contours correspond to 10$\sigma$, 20$\sigma$, 50$\sigma$, 75$\sigma$, 
100$\sigma$, and 150$\sigma$ above the background level.  
}
\label{ximg}
\end{figure*}

To study the spatial distribution of the X-ray emission from NGC\,3242, 
we have produced EPIC images of NGC\,3242 in the energy band 0.25-2.5 
keV by extracting the individual EPIC-pn, EPIC-MOS1, and EPIC-MOS2 
images, mosaicing them together, applying the exposure map correction, 
and smoothing it.  
The raw image is shown in the left panel of Figure~\ref{ximg}, while 
the exposure map corrected, smoothed version of the image overlaid by 
X-ray contours is shown in the central panel. 

To compare the relative spatial distribution of the X-ray-emitting 
gas and the ionized nebular material, we have superposed the X-ray 
contours on the \emph{HST} WFPC2 [O~{\sc iii}] image of NGC\,3242 
(Fig.~\ref{ximg}-{\it right}).
The comparison between the X-ray contours and the optical image suggests 
that the X-ray-emitting gas is confined within the innermost shell of 
NGC\,3242, as the location of the lowest intensity level X-ray contours 
outside the 28\arcsec$\times$20\arcsec\ inner shell of NGC\,3242 is most
likely caused by the point-spread function (PSF) of \emph{XMM-Newton} 
that is known to have a half energy width (HEW) of 15\farcs2 for 
EPIC-pn and 13\farcs0-13\farcs8 for EPIC-MOS and significantly extended 
wings.

\begin{figure*}[Ht!]
\begin{center}
\includegraphics[bb=20 275 590 535,width=14.0cm]{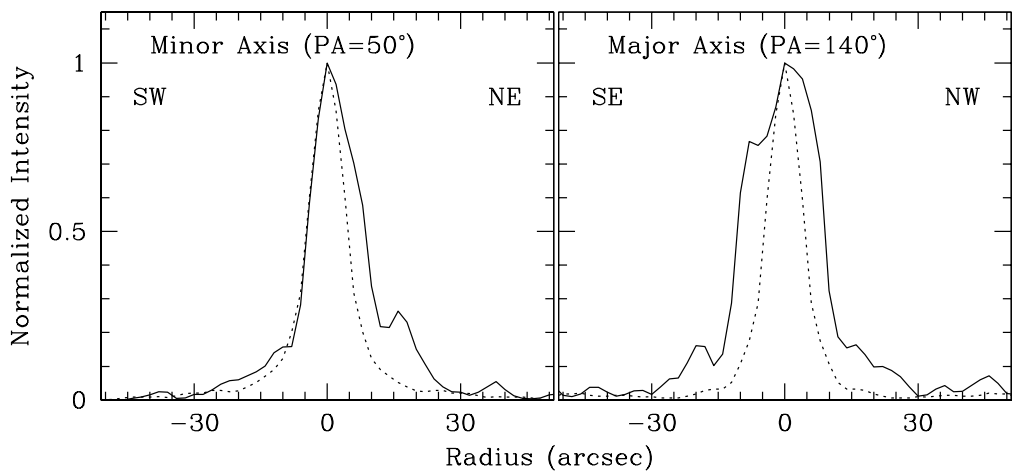}
\end{center}
\caption{
Normalized intensity of the EPIC-MOS X-ray surface brightness profile 
of NGC\,3242 (solid line) and a point-source in the field of view of 
the instrument (dotted line) along the minor ({\it left}) and major 
({\it right}) axes of the nebula.  
}
\label{prof}
\end{figure*}

The X-ray image and contours also suggest that the detailed morphology 
of the X-ray emission is asymmetric, with an emission peak northwest of the 
central star.  
%We will discuss it later, but it seems unlikely that the asymmetrical 
%X-ray morphology of NGC\,3242 could be due to uneven internal extinction 
%since the extinction towards NGC\,3242 is very low.  
%Note, however, that the brightest X-ray emission is found in the 
%region of the nebula that is moving towards us, (i.e., that is 
%close to us) and that could be affected by a smaller hydrogen column 
%density.
We further compare in Figure~\ref{prof} the surface brightness profile 
of the X-ray emission of NGC\,3242 along the major and minor nebular 
axes with the PSF from a point source in the field of view.  
Both profiles of the nebular emission are asymmetric and more 
extended than those of the point-source, more clearly for the 
profile along the major axis of the nebula. 

In order to better understand the distribution of X-ray-emitting gas in 
NGC\,3242 we have constructed simulated observations for comparison with 
the \emph{XMM-Newton} images.  
For these simulated observations we modeled the emission from NGC\,3242 
as though it arose from a constant-density ellipsoidal shell of 
X-ray-emitting gas interior to the innermost nebular shell.  
We assumed a prolate ellipsoid with semi-major and semi-minor
axes of 11\farcs25 $\times$ 7\farcs5 in the plane of the sky.  
Three models were considered: 
Model A assumed the X-ray-emitting gas was confined to an ellipsoidal shell
with a fractional width ($\Delta r/r$) of 0.1; Model B assumed a 
similar shell with a fractional width of 0.2; and Model C assumed the 
X-ray-emitting gas filled the ellipsoid with constant density.

\begin{figure*}[Ht!]
\begin{center}
\includegraphics[bb=40 265 485 424,width=15.0cm]{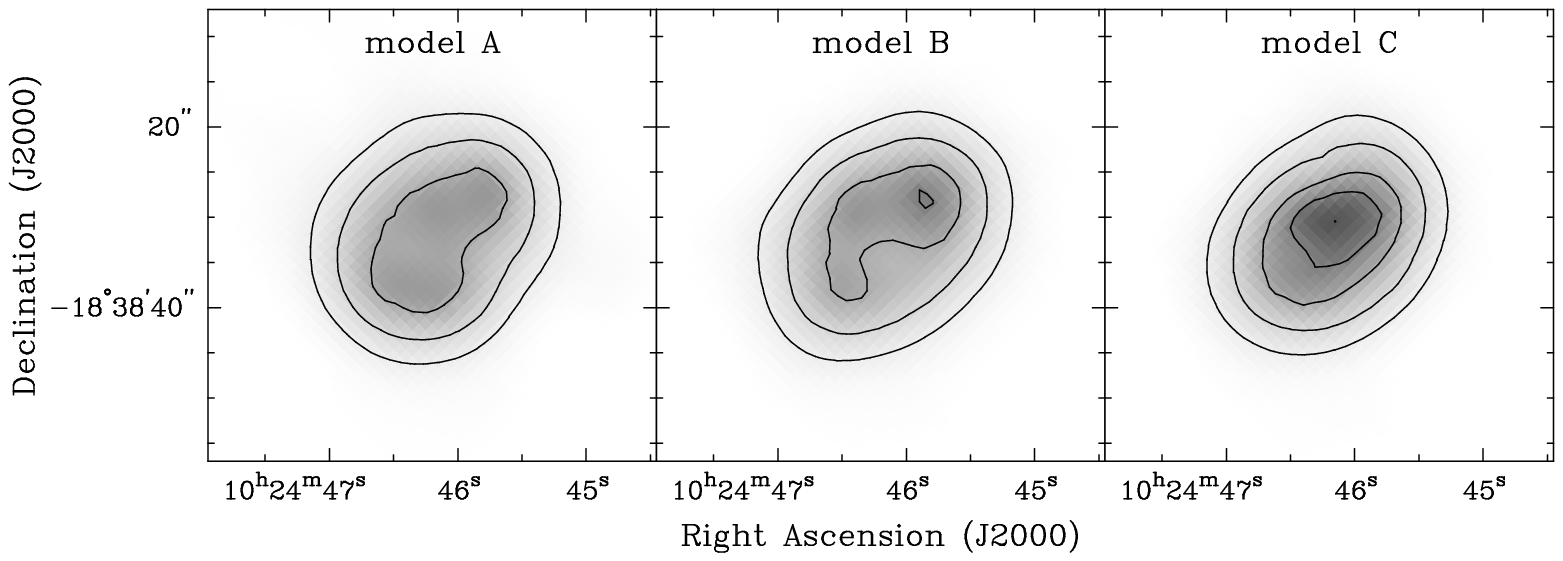}
\end{center}
\caption{
Realizations of three Monte Carlo simulations of the \emph{XMM-Newton} 
images of NGC\,3242 for an X-ray-emitting ellipsoidal shell of constant 
density with a shell thickness 10\% (model A) and 20\% (model C), and 
a filled shell (model C).  
Contours correspond to 20$\sigma$, 40$\sigma$, 80$\sigma$, and 
120$\sigma$ above the background level.
}
\label{sim_ximg}
\end{figure*}

To simulate yhe \emph{XMM-Newton} observations, we generated random simulated 
X-rays based on the model distribution of the X-ray emitting gas with further 
randomization in the plane of the sky consistent with the \emph{XMM-Newton} 
PSF\footnote{
Based on \emph{XMM-Newton} observations of the bright, soft point source 
Nova LMC1995 \citep{Orio_etal2003}.}.  
We also added random X-ray events to the simulated image to mimic
the background emission. The total number of simulated source and noise counts 
were set to match those of the actual \emph{XMM-Newton} observations.  
Each model was used to make ten simulations to explore the variations
caused by the small number statistics of these Monte Carlo simulations.
We then adaptively smoothed the simulated observations with the same parameters
used for the actual observations.  
Figure~\ref{sim_ximg} shows a typical realization for each of 
the three models described.  In the 
realizations of models A and B (the shell models) a central deficit in 
X-ray emission is always apparent.  In the realizations of Model C the 
X-ray emission peaks on or near the center; the example shown in 
Fig.~\ref{sim_ximg} even exhibited an offset peak matching the 
actual observation.  
These simulated observations demonstrate that the diffuse X-ray emission 
from NGC\,3242 is more consistent with that from a central cavity filled 
with X-ray-emitting gas than from a thin ellipsoidal shell. 
They also suggest that the asymmetric distribution of the X-ray 
emission may be spurius due to the low count number.

\subsection{Spectral Properties of the X-ray Emission from NGC\,3242}

\begin{figure}[Ht!]
 \begin{center}
\includegraphics[angle=360,width=7.5cm]{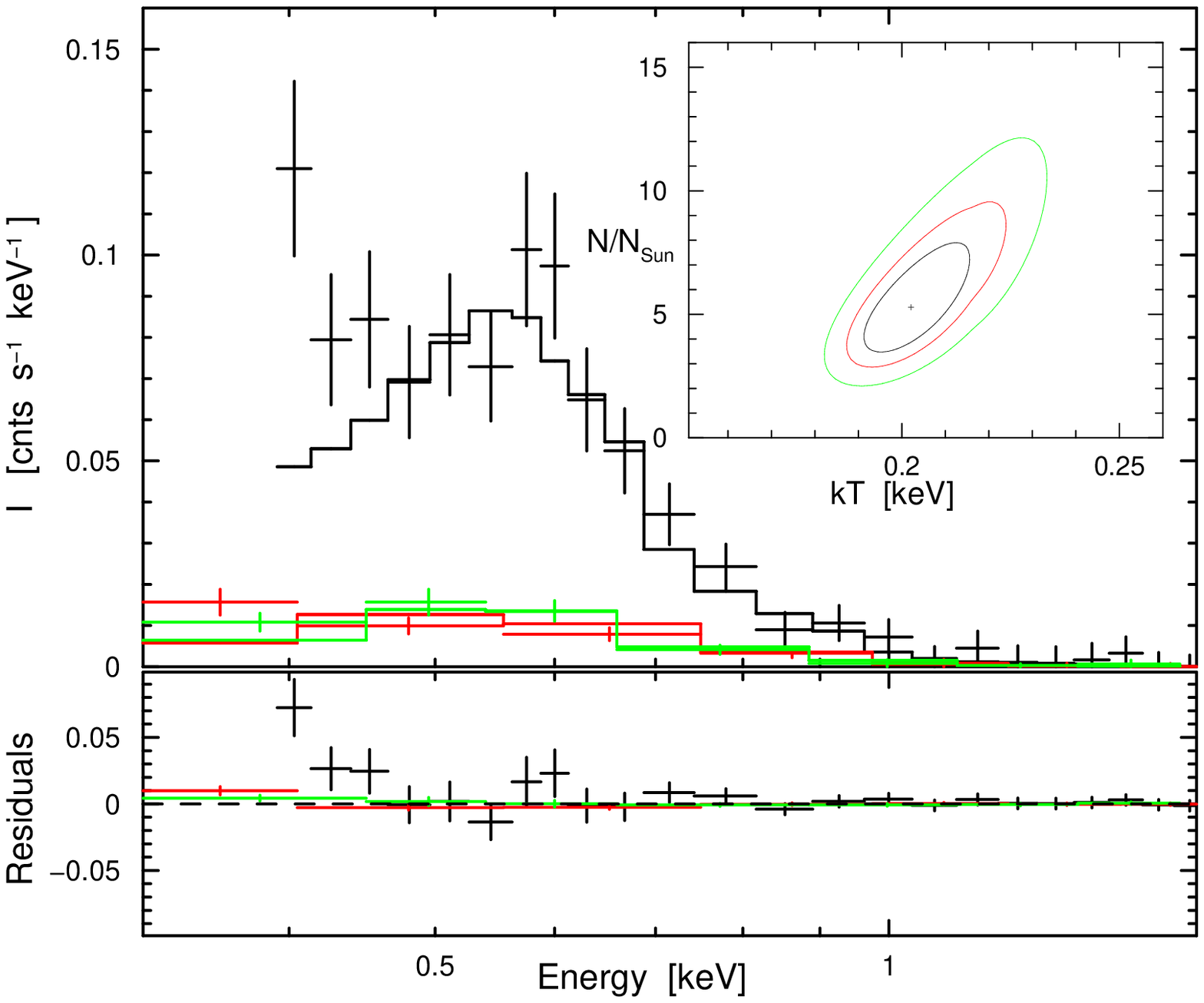}
 \end{center}
 \caption{({\it top-panel}) 
EPIC-pn (black), EPIC-MOS1 (red), and EPIC-MOS2 (green) background-subtracted 
spectra of the diffuse emission of NGC\,3242.  
The best-fit joint model is shown as a histogram in the corresponding 
color.  
The inset shows the nitrogen vs.\ temperature $\chi^{2}$ grid plot of the 
spectral fit where the black, red, and green curves represent the 68\%, 
90\%, and 99\% confidence levels.  
({\it bottom-panel}) 
Residuals of the best-fit joint model to the EPIC-pn (black), EPIC-MOS1 
(red), and EPIC-MOS2 (green) spectra of NGC\,3242 shown in the top panel.}
 \label{xspec}
\end{figure}

To study the spectral properties of the X-ray emission from 
NGC\,3242, we have extracted its EPIC-pn, EPIC-MOS1, and 
EPIC-MOS2 background-subtracted spectra (Figure~\ref{xspec}).  
Our description of the spectral properties of NGC\,3242 will 
focus on the EPIC-pn spectrum as the number of counts in this 
spectrum is $\sim$5 times larger than in the EPIC-MOS1 and 
EPIC-MOS2 spectra.  
The EPIC-pn spectrum of NGC\,3242 is soft, with most counts 
below 1.0 keV.  
The spectrum peaks at 0.5-0.6 keV, and then steadily declines towards 
higher energies. %with a subtle emission peak at $\sim$0.9 keV.  
This peak is most likely due to the He-like O~{\sc vii} triplet
at 0.57 keV.
%He-like O~{\sc vii} and Ne~{\sc ix} triplets at 0.57 keV and 0.92 keV

We shall note that the shape of the EPIC-pn spectrum at energies 
below 0.55 keV is difficult to explain.  
The unbinned spectrum reveals noticeable oscillations in the 
count rates with small number of counts in energy bins at 
$\sim$450 eV and $\sim$500 eV, and large number of counts in 
energy bins between these two energies and at $\sim$400 eV.  
Since these ``spectral features'' have widths smaller than the EPIC-pn 
spectral resolution at this energy range (FWHM$\sim$90 eV), it is unlikely 
that these spikes are associated with real emission lines.  
It has been reported that optical loading can produce the observed 
effects, but this mechanism can be ruled out because the Medium Filter 
used in our observations prevents the optical contamination from 
point sources as bright as $m_V$=6--9 mag.\ (XMM-SOC-CAL-TN-0051), 
while the central star of NGC\,3242 has $m_V$=10.3 mag \citep{vALH95}.  
We can conclude that these oscillations are most likely caused 
by stochastic effects.  
To mitigate this issue, the EPIC-pn spectrum has been binned to have at 
least 25 counts per channel for further spectral analysis.  
Similarly, the EPIC-MOS spectra have been binned to have at 
least 15 counts per channel for spectral analyses.

\subsection{Spectral Analysis}

For the spectral analysis, we have adopted the nebular chemical abundances 
(He=0.94\,He$_\odot$, C=0.78\,C$_\odot$, N=1.61\,N$_\odot$, O=0.83\,O$_\odot$, 
Ne=0.75\,Ne$_\odot$, S=0.20\,S$_\odot$, and Ar=0.40\,Ar$_\odot$), and 
foreground hydrogen column density ($N_{\rm H}$=5$\times$10$^{20}$ cm$^{-2}$) 
derived by \citet{POT08}.  
We have then modeled the observed EPIC spectra using an absorbed 
APEC optically thin plasma emission model and adopting the absorption 
cross-sections from \citet{MMcC83}.  

% A photoelectric absorption using WABS \textit{Wisconsin 
% (Morrison and McCammon)} cross-sections.\[A(E) = e^{(-n_{H}*\sigma(E))}\]
% Where $\sigma(E)$ is the photo-electric cross-section (NOT including 
% Thomson scattering) and $n_{H}$ equivalent hydrogen column (in units of 
% $10^{22}$ atoms/$cm^{2}$)

\begin{figure}[Ht!]
 \begin{center}  
\includegraphics[angle=270,width=7.5cm]{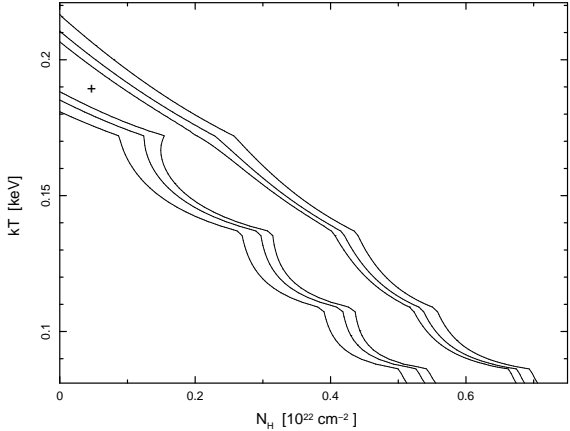}
 \end{center}
 \caption{   
Temperature vs.\ column density $\chi^{2}$ grid plot of the
spectral fit with nebular abundances and free column density,
where the three curves represent the 68\%, 90\%, and 99\%
confidence levels, and the cross marks the best fit for the
value of 5$\times$10$^{20}$ cm$^{-2}$ adopted for $N_{\rm H}$.
}
 \label{TNH_cor}
\end{figure}

This model provides a reasonable fit to the EPIC spectra of NGC\,3242's 
diffuse emission with a reduced $\chi^{2}$ of 1.60 (=38.4/24) for 
$kT$=0.190$\pm$0.009 keV ($\sim$2.2$\times$10$^6$~K), although the 
best-fit model is lower than the observed spectrum at energies $<$0.5 keV.
The fits are not improved by allowing $N_{\rm H}$ to vary and we find
that the best-fit values of $kT$ and $N_{\rm H}$ appear to be correlated 
(Figure~\ref{TNH_cor}) as $kT = 0.1975 - 20.0{\times}N_{\rm H}$, where 
$kT$ is given in keV and $N_{\rm H}$ in units of 10$^{20}$ cm$^{-2}$. 

Alternatively, we may allow the chemical abundances of nitrogen and carbon 
to vary in the fits, as these two elements have spectral lines that can 
contribute in the 0.4-0.5 keV energy band.  
Variations of the carbon abundance do not produce noticeable changes 
in the quality of the spectral fit, and thus the carbon abundance cannot
be well constrained.
On the other hand, changes to the nitrogen abundances produce a 
significant improvement of the spectral fit, and the
best-fit model (Figure~\ref{xspec}) has $kT$=0.202$^{+0.012}_{-0.010}$ keV 
($\sim$2.35$\times$10$^6$~K), N=5.3$^{+2.8}_{-1.8}$\,N$_\odot$, and
a reduced $\chi^{2}$ of 1.07 (=24.56/23).
The limits to the nitrogen abundance set by this fit are better
illustrated by the nitrogen versus temperature $\chi^{2}$ grid plot 
of the spectral fit in Figure~\ref{xspec}, which shows
that the range of nitrogen abundances exclude the nebular
abundance of 1.61\,N$_\odot$.  
The best-fit value for the nitrogen abundance implies a N/O ratio 
of the X-ray-emitting gas to be (N/O)$_{\rm X}\sim$1.2, $\sim$3.3 
times greater than the N/O ratio of the ionized nebular material, 
(N/O)$_{\rm neb}\sim$0.36.

For the best-fit model with enhanced nitrogen abundance, we 
derive an observed flux of 
(4.2$^{+0.7}_{-1.1}$)$\times$10$^{-14}$ ergs~cm$^{-2}$~s$^{-1}$
and an intrinsic X-ray luminosity of
(7.3$^{+0.8}_{-1.3}$)$\times$10$^{30}\,d^2$ ergs~s$^{-1}$
in the 0.4-2.0 keV energy band, where $d$ is the distance in kpc.  
The volume emission measure ($EM=N_{\rm e}^2\,V$, where $N_{\rm e}$
is the electron density and $V$ is the emitting volume) of this
model is 
% 
% K * 4 * PI * D^2 / 1.0e-14 
% 
$\sim$5$\times$10$^{53}\,d^2$ cm$^{-3}$.
The emitting volume of the inner shell of NGC\,3242 is
$\sim$2$\times$10$^{52}\,\epsilon\,d^3$ cm$^3$, 
where $\epsilon$ is the volume filling factor of the 
X-ray-emitting gas.
Thus, the electron density of the X-ray-emitting gas is
$\sim$5\,$\epsilon^{-1/2}\,d^{1/2}$ cm$^{-3}$, and
the thermal pressure is 
1.6$\times$10$^{-9}\,\epsilon^{-1/2}\,d^{1/2}$ 
dyne~cm$^{-2}$.  
At a distance of 0.55 kpc, the X-ray luminosity of NGC\,3242 is 
(2.2$^{+0.2}_{-0.4}$)$\times$10$^{30}$ ergs~s$^{-1}$, the electron 
density of the X-ray-emitting gas is 4\,$\epsilon^{-1/2}$ cm$^{-3}$, 
and its thermal pressure is 1.2$\times$10$^{-9}\,\epsilon^{-1/2}$ 
dyne~cm$^{-2}$.

\section{Physical Structure of the Optical Shell}

\subsection{Bulk Physical Conditions}

The electron temperature and density of the optical shell of NGC\,3242 can 
be derived using standard techniques \citep[e.g.,][]{GMS-R1996} from the 
long-slit intermediate-dispersion spectroscopy presented in \S2.3.  
From these two-dimensional spectra of NGC\,3242, we have extracted 
one-dimensional spectra of representative regions of its inner and 
outer shells, and measured the fluxes of the emission lines using 
the IRAF task \emph{splot}.  Table~\ref{tab.flux} presents line fluxes
normalized to an H$\beta$ flux of 100.
The observed fluxes, $F$, have been dereddened using the IRAF task 
\emph{redcorr} to derive the intrinsic intensity of the line, $I$:  
\begin{equation}
I=F \times 10^{-c_{H\beta} \times f_\lambda}
\end{equation}
where c$_{H\beta}$ is the logarithmic H$\beta$ extinction constant 
computed by comparing the observed value of the H$\alpha$ to H$\beta$ 
ratio to the expected theoretical value of 2.87 for case B 
recombination \citep{OST06}.  
The observed fluxes are subsequently corrected using the values of 
$f_\lambda$ corresponding to the interstellar extinction law of \citet{SAV79}.  
The H$\alpha$/H$\beta$ ratio measured in the medium dispersion spectra 
implies an extinction coefficient c$_{H\beta}$=0.01 for the inner shell 
and 0.05 for the outer shell.
We note that these values of c$_{H\beta}$ are smaller than the values 
0.10-0.15 derived by \citet{Balick_etal93} and \citet{POT08}, but consistent 
with the values derived by \citet{HE00} using this same dataset.

The physical conditions for the inner and outer shells of NGC\,3242 have 
been derived using  temperature-sensitive line ratios of [O~{\sc iii}], 
[N~{\sc ii}], [O~{\sc ii}], and [S~{\sc iii}], and density-sensitive 
line ratios of [S~{\sc ii}] and [Ar~{\sc iv}].  
The ions and lines used, their ratios, and the values of electron 
temperature, $T_{\rm e}$, and density, $N_{\rm e}$, are listed in 
Table~\ref{tab.tn}. 
The results suggest that the inner shell density is $\sim$2,200~cm$^{-3}$, 
in excellent agreement with the density derived from the [S~{\sc ii}] lines 
by \citet{POT08}, and the outer shell density is $\sim$370~cm$^{-3}$.
Among the four temperatures derived from different lines, we adopt that
from the [O~{\sc iii}] lines, 11,900~K for the inner shell and 10,400~K 
for the outer shell, as the [O~{\sc iii}] temperature diagnostic is the 
least affected by the extinction law and the lines are the brightest.
This radial decline in temperature is consistent with the results 
presented by \citet{Balick_etal93}.  
Using these values, we derive thermal pressures of 3.6$\times$10$^{-9}$ 
dyne~cm$^{-2}$ and 5$\times$10$^{-10}$ dyne~cm$^{-2}$ for the inner and 
outer shells, respectively.

\subsection{Spatially Resolved Physical Conditions}

\begin{figure*}[Ht!]
\begin{center}
\includegraphics[angle=360,width=15cm]{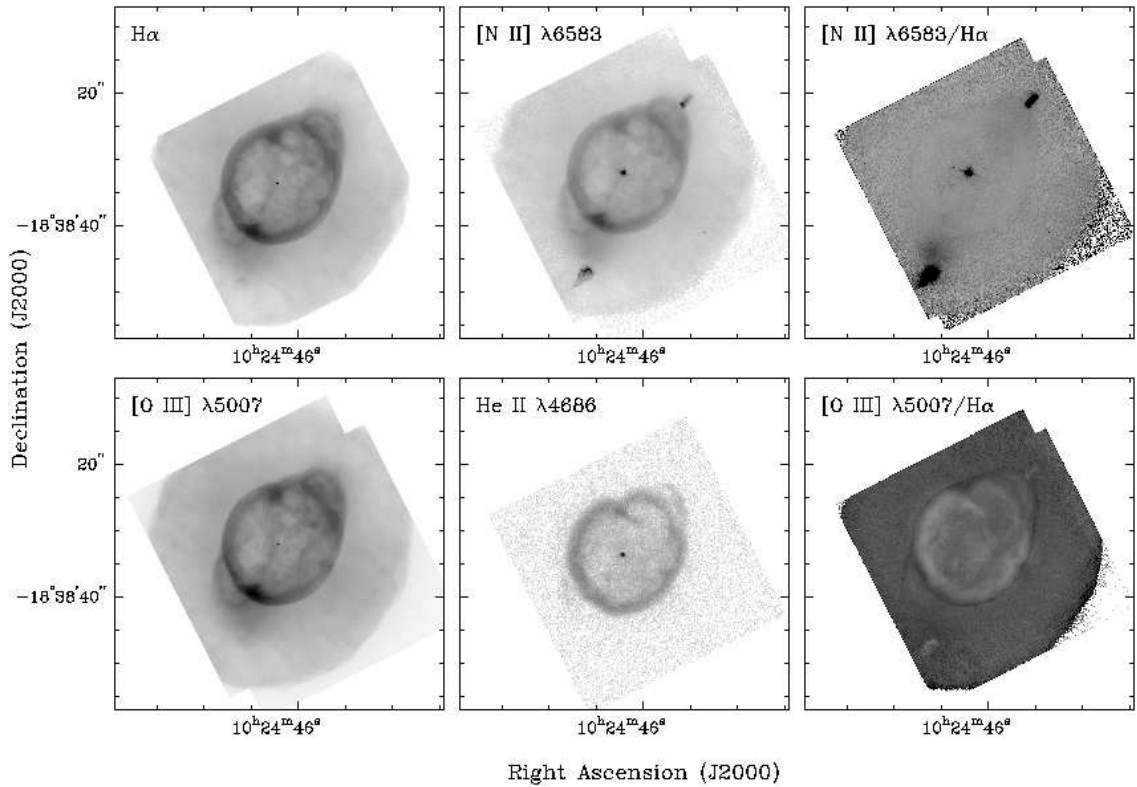}
\end{center}
\caption{
\emph{HST} WFPC2 narrow-band images and ratio maps of the central region 
of NGC\,3242.  
The images are displayed on a squared-root scale, while the ratio 
maps are displayed on a linear scale.
}
\label{hst_img}
\end{figure*}

The narrow-band \emph{HST} images and ratio maps of NGC\,3242 in 
different emission lines (Figure~\ref{hst_img}) display a simple
shell morphology consisting of an elliptical inner shell and an
attached outer shell.  This shell morphology is only moderately 
complicated by ansae and knots best seen in the [N~{\sc ii}] 
images \citep[the so-called Fast Low-Ionization Emission Regions, 
FLIERS,][]{Balick_etal98}.
A spatio-kinematic study of NGC\,3242 by \citet{BPI87} confirms 
this simple shell structure: the inner shell is a nearly round 
bubble expanding at 25--30 km~s$^{-1}$, while the outer shell 
is a co-expanding envelope filled with material. 
Thus, the H$\alpha$ image of NGC\,3242 can be used to obtain
density and pressure profiles of the nebula for comparison with
those of the X-ray-emitting gas in the central cavity.

\begin{figure}[Ht!]
 \begin{center}
\includegraphics[bb=45 235 555 610,width=7.5cm]{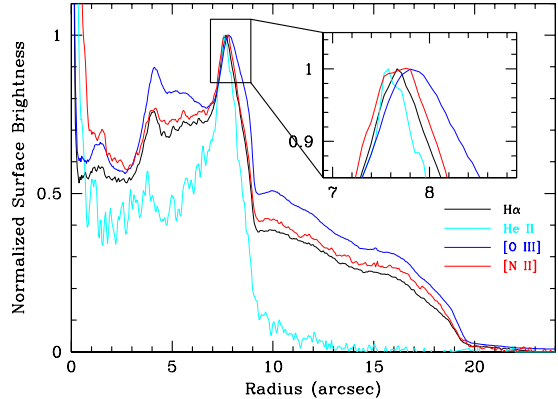}
 \end{center}
 \caption{
H$\alpha$, He~{\sc ii} $\lambda$4686, [O~{\sc iii}] $\lambda$5007, 
and [N~{\sc ii}] $\lambda$6583 surface brightness profiles of the 
inner and outer shells of NGC\,3242 normalized to the bright peak 
at the rim of the inner shell.  
The profiles are derived by averaging the surface brightness in radial
bins in a sector along PA = 230\degr\ with $\Delta$PA = 10\degr.
The inset expands this peak to better illustrate the spatial 
differences at this location of the emission in the different 
lines.  
}
\label{fig:prof_obs}
\end{figure}

The density profile of NGC\,3242 has been previously determined
by \citet{SZB92} using a ground-based H$\alpha$ image.  We will
use the high-resolution \emph{HST} WFPC2 H$\alpha$ image to determine
the electron density, $N_{\rm e}$, and thermal pressure, $P_{\rm th}$, 
of the inner shell and the envelope as a function of nebular radius.
First, we examined this image to select a direction that 
would provide us with a clean surface brightness profile.  
The cut along PA=230\arcdeg\ is orthogonal to the inner shell rim 
and covers the full extent of the outer shell.  Thus, we extracted 
the surface brightness profiles in the H$\alpha$, He~{\sc ii} 
$\lambda$4686, [O~{\sc iii}] $\lambda$5007, and [N~{\sc ii}] 
$\lambda$6583 shown in Figure~\ref{fig:prof_obs}.  
The surface brightness profile of the outer shell is very similar 
in the H$\alpha$, [O~{\sc iii}], and [N~{\sc ii}] lines, but the 
He~{\sc ii} emission is diminished most likely because the inner 
shell is optically thick to He$^+$ ionizing photons.  
As for the inner shell, all profiles show the bright shell rim, 
but peak at different radial distances: 
$\sim$7\farcs4 for He~{\sc ii}, 
$\sim$7\farcs6 for H$\alpha$, and 
$\sim$7\farcs8 for [O~{\sc iii}].  
The emission inside this rim drops steeply, but the filaments
projected in the central cavity produce secondary peaks in the
H$\alpha$, [O~{\sc iii}], and [N~{\sc ii}] lines, but not in 
the He~{\sc ii} line.  

\begin{figure*}[Ht!]
 \begin{center}
\includegraphics[bb=45 235 555 610,width=8cm]{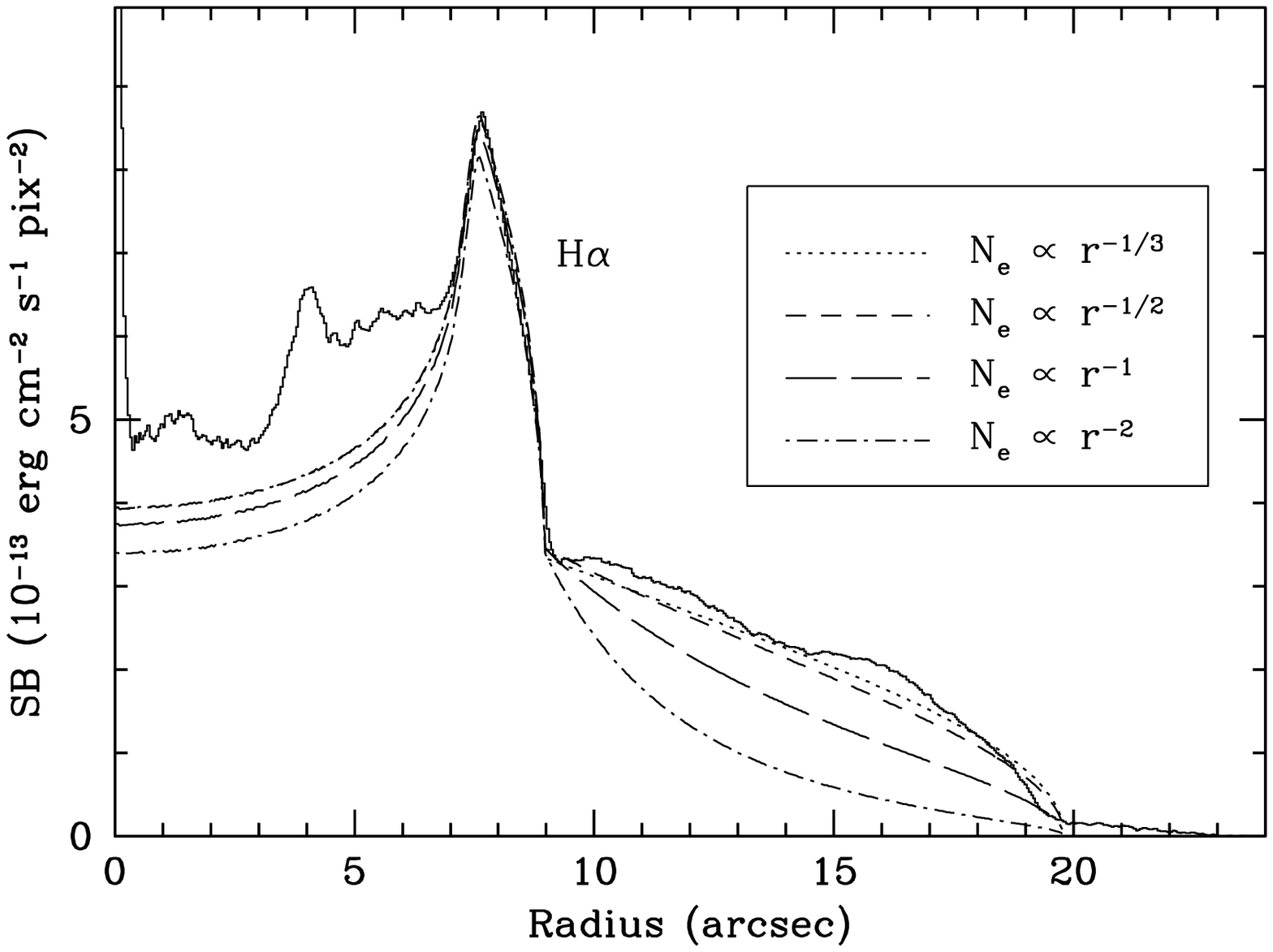}
\includegraphics[bb=45 235 555 610,width=8cm]{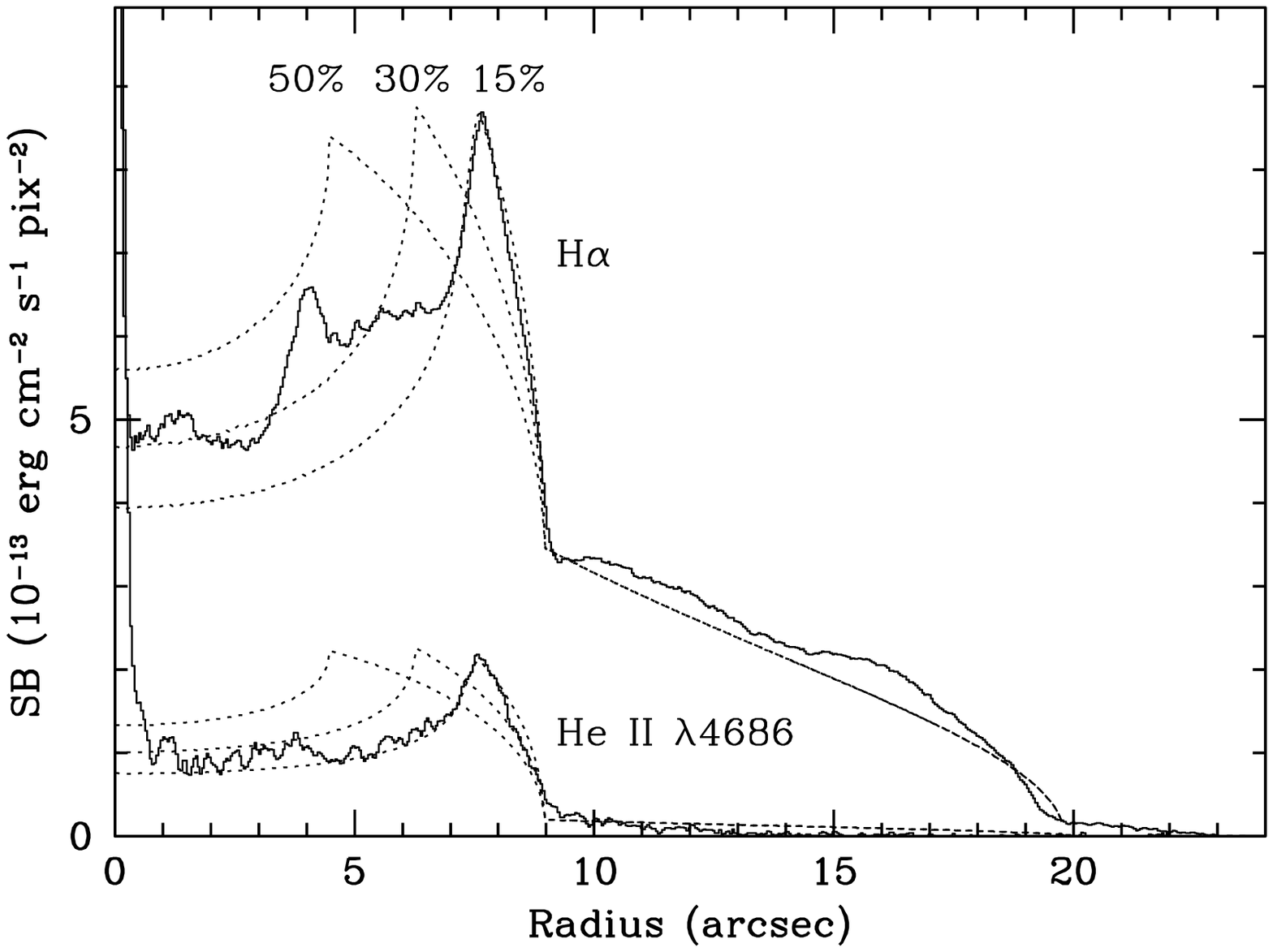}
 \end{center}
 \caption{
({\it left}) 
H$\alpha$ surface brightness profile of NGC\,3242 along PA=230$^\circ$ 
(thick line) and synthetic H$\alpha$ surface brightness profiles of the 
outer shell ($r>$8\farcs9) for different radial dependences of $N_{\rm e}$.  
As for the inner shell, a thickness of 15\% in radius has been assumed (see 
below).  
({\it right}) H$\alpha$ and He~{\sc ii} $\lambda$4686 surface brightness 
profile of NGC\,3242 along PA=230$^\circ$ (thick lines) and synthetic 
surface brightness profiles of the inner shell for different values of 
the shell thickness: 15\%, 30\%, and 50\%.  
As for the outer shell, a decay of $N_{\rm e}{\propto}r^{-1/2}$ 
has been assumed (see above).  
}
\label{prof:ne}
\end{figure*}

The outer shell has been modeled assuming a mean electron density 
$\sim$370~cm$^{-3}$ and a temperature of 10,400~K.  
Figure~\ref{prof:ne}-{\it left} shows the fit to the H$\alpha$ surface 
brightness profile of the outer shell of NGC\,3242 assuming four
different radial dependences of the density:
$ r^{-2}$, $r^{-1}$, $r^{-1/2}$, and $r^{-1/3}$.
The model surface brightnesses for the $ r^{-2}$ and $r^{-1}$ density
profiles decrease outwards too rapidly compared to the observed
surface brightness profile, and can thus be excluded.
The model surface brightness for the $r^{-1/2}$ and $r^{-1/3}$ density
profiles more closely match the observation.

Using the $r^{-1/2}$ density profiles in the outer shell,
we proceed to model the 
H$\alpha$ and He~{\sc ii} surface brightness profiles at the 
rim of the inner shell by varying the thickness of the shell 
and adjusting the ionization fraction of He$^{++}$.  
A shell with a constant density 2,200~cm$^{-3}$, temperature of 11,900~K, 
shell thickness $\sim$15\%, and a sharp decay in its inner edge produces 
a good fit to the He~{\sc ii} profile.  
The fit to the rim of the shell in the H$\alpha$ profile is also 
reasonable, but the model profile does not include the contributions from 
filaments to the H$\alpha$ profile (Figure~\ref{prof:ne}-{\it right}).

\begin{figure*}[Ht!]
\begin{center}
\includegraphics[bb=30 280 590 490,width=16.5cm]{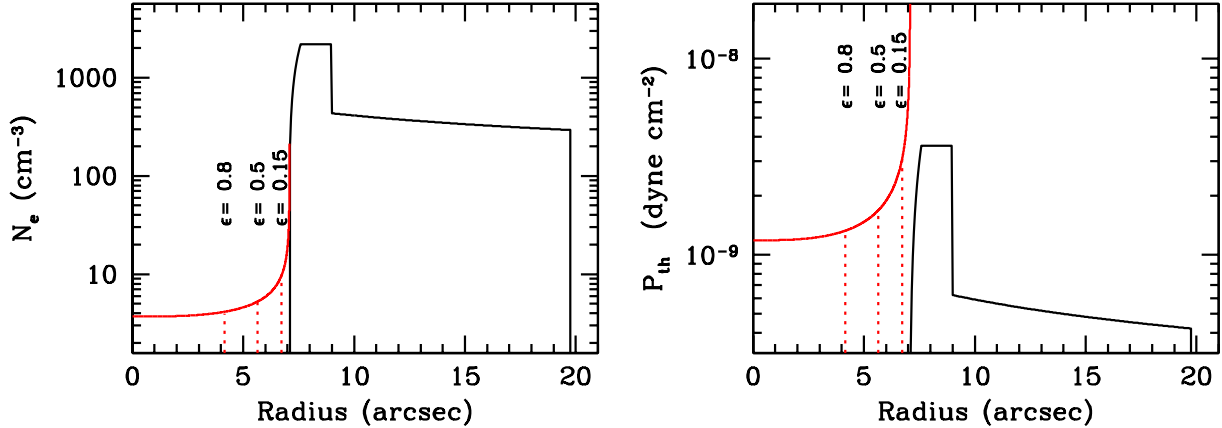}
\end{center}
\caption{
({\it left}) Electron density, $N_{\rm e}$, and ({\it right}) thermal 
pressure, $P_{\rm th}$, radial profiles of the central cavity
and the inner shell and envelope of NGC\,3242.  
The density and thermal pressure of the X-ray-emitting gas within the 
central cavity (radius $\lesssim$ 7\arcsec) are shown as a function of 
the filling factor, $\epsilon$ {\bf (red curves)}.  
The values of these physical conditions for filling factors, $\epsilon$, 
of 0.15, 0.5, and 0.8 are further labeled on the plot at the 
inner radii of the corresponding hypothetical hot gas shell.  
}
\label{prof:phys}
\end{figure*}

The density profile deduced from the model comparisons is shown in 
Figure~\ref{prof:phys}-{\it left}.  
The radial profile of the thermal pressure (Figure~\ref{prof:phys}-{\it right}) 
has been constructed assuming a constant electron temperature of 11,900~K for 
the inner shell, and 10,400~K for the outer shell, as derived from temperature 
sensitive optical line ratios.  
This figure shows that the thermal pressure of the inner shell is 
much higher than that of the outer shell, i.e., the inner shell 
is expanding into the envelope.

We can also compare the thermal pressure of the nebular shell 
to that of the hot interior.  
Assuming that the hot gas is distributed in a constant-density shell 
interior to but in contact with the inner nebular shell, the filling 
factor can be expressed as $\epsilon =1-(r_i/r_o)^3$, where $r_o$ and 
$r_i$ are the outer and inner radii of the hot gas shell ($r_o$ is 
also the inner radius of the inner nebular shell).
As the hot gas density is $\sim$4\,$\epsilon^{-1/2}$ cm$^{-3}$ ($d=0.55$ kpc), 
we have plotted the hot gas density as a function of its inner shell radius 
$r_i$ in Figure~\ref{prof:phys}-{\it left} and the thermal pressure in 
Figure~\ref{prof:phys}-{\it right}.  
The hot gas shell's inner radii corresponding to $\epsilon$ of 
0.15, 0.5, and 0.8 are marked.
It is apparent that the hot gas pressure exceeds that of the nebular
shell only for small filling factor values, $\epsilon < $0.15, thus 
requiring the hot gas to be concentrated in a thin shell with fractional
thickness $\lesssim$0.05.

\section{Discussion}

The properties of the X-ray emission from NGC\,3242, based on a 
preliminary analysis \citep{Ruiz_etal06}, were compared to 1D 
hydrodynamical models of PNe that included the effects of stellar 
wind evolution and heat conduction by \citet{SSW08}.  
They remarked that the round shape of the rim of the inner 
shell of NGC\,3242 makes it a suitable PN for the comparison 
with their 1D simulations.  
The revised values of different X-ray properties of NGC\,3242 presented 
here need to be discussed in the framework of \citet{SSW08}'s model.  
This revision mostly affects the values of the thermal 
pressure of the hot bubble and inner shell rim, whereas 
the X-ray luminosity (scaled to the distance of 0.55 kpc) 
and plasma temperature of NGC\,3242 are basically the same 
as those used by \citet{SSW08}.

We note that the models selected by \citet{SSW08} have a thermal 
pressure of the hot bubble that exceeds that of the rim, while 
our revised values indicate the opposite, i.e., the thermal pressure 
of the rim ($P_{\rm th}$=3.6$\times$10$^{-9}$ dyne~cm$^{-2}$) is 
higher than that of the bubble 
($P_{\rm th}$=1.6$\times$10$^{-9}\,\epsilon^{-1/2}\,d^{1/2}$ 
dyne~cm$^{-2}$).  
One possible outcome to this issue would be a distance for NGC\,3242 
much farther than 0.55 kpc \citep{TZ97,ML04}; at a distance of 3 kpc, 
the thermal pressures of the hot bubble and rim would be the same, but 
this would make NGC\,3242 too large ($r=0.12$ pc) and evolved ($\tau=6,000$ 
yrs) and we consider it to be unlikely.

The other possible solution is to consider a low value for the filling 
factor of the X-ray-emitting gas in the bubble; a value $\epsilon$=0.15 
would make both pressures the same.  
Such a low filling factor is expected to produce a noticeable 
limb-brightening morphology that is not observed 
(Figures~\ref{ximg} and \ref{prof}).
As shown in Fig.~\ref{sim_ximg}, hints of a limb-brightened morphology
are still apparent for a hot gas shell with a fractional thickness of 
0.2 (i.e., $\epsilon \sim 0.5$) even at the \emph{XMM-Newton} limited 
spatial resolution ($\sim15^{\prime\prime}$).
It is arguable, however, that extinction has reduced the 
center-to-limb contrast of the X-ray emission, thus shifting 
the emission inwards, as this is an effect expected to be 
noticeable \citep{SSW08} even for the small extinction 
($N_{\rm H}$=5$\times$10$^{20}$ cm$^{-2}$) towards the nebula.

Alternatively, we must consider the dynamical effects of the 
photo-ionization of the nebular material, which may have
overcome those of the currently diminished stellar wind of 
NGC\,3242 \citep{Kudritzki_etal97,TL02}.  
In this case, the nebular expansion is driven by 
the thermal pressure increase produced in the rim 
by photo-ionization.  
This may explain the relatively large thickness of NGC\,3242 inner 
shell rim, as compared to that of NGC\,6543.  
We suggest that the rim thickness can be used as an indicator of the 
relative thermal pressure of hot gas and photo-ionized nebular material;  
a sharp rim would imply that the thermal pressure of the hot gas is 
larger than this of the nebula, and thus that hot gas itself is present 
(and detectable), while a thick inner shell rim would imply a lower 
pressure for the hot gas, if present.

The low plasma temperature is puzzling in view of the enhanced N/O 
ratio of the X-ray-emitting plasma, about three times higher than 
the N/O ratio of the nebular material.  
Such differences in the chemical abundances seem to imply that the 
X-ray-emitting plasma mostly consists of shocked stellar wind, with 
little contamination of material from the nebula.  
The low plasma temperature, however, needs large amounts of 
material from the cold nebular shell to have been incorporated 
into the hot bubble.  
This problem, similar to that found in other PNe such as 
BD+30\degr3639 \citep{Yu_etal2009}, NGC\,2392 \citep{GR05}, and 
NGC\,6543 \citep{CH01}, may require mechanisms for the production 
of X-ray emission in PNe other than mixing and heat conduction as 
summarized by \citet{Soker_etal2010}.
% This is a problem similar to that found in other PNe 
% such as NGC\,6543 \citep{CH01}.  

There is, finally, a noticeable asymmetry of the spatial distribution 
of the X-ray emission from NGC\,3242, with the northwestern half of the 
hot bubble being brighter than its southeastern half (Figures~\ref{ximg} 
and \ref{prof}).  
Even if we assume that the inner shell of NGC\,3242 were a tilted 
ellipsoid with its Northwestern tip moving towards us (and its 
Southeastern tip receding from us), it seems unlikely that an uneven 
spatial distribution of intervening material could be responsible of 
the observed asymmetry as it may be the case of the dusty PNe 
BD+30$^\circ$3639 \citep{KN00} and NGC\,7027 \citep{KN01}, because 
only small extinction variations were observed in NGC\,3242 
\citep{Balick_etal93}.  
On the other hand, our simulations in Section 3.1 suggest 
that the asymmetric distribution of X-ray emission could 
result from small number statistics.  

%the axis of symmetry of the X-ray emission may be not 
%completely coincident with the inner shell axis of symmetry, and it 
%is definitely not aligned with the FLIERs.  

% \subsection{Complete physical structure of NGC\,3242 }

% To investigate the physical structure of NGC\,3242, we have combined 
% these X-ray observations with narrow-band \emph{HST} images and 
% intermediate dispersion spectra.  
% The high-dispersion spectra indicate that the inner shell can 
% be described as a tilted ellipsoidal shell with an equatorial expansion 
% velocity of 26 km~s$^{-1}$.  
% At a distance of $\sim$1.0 kpc, as derived from a study of the 
% expansion of NGC\,3242 using multi-epoch \emph{HST} images, its 
% kinematic age is $\sim$1,700 yrs.  

\section{Summary}

We have obtained \emph{XMM-Newton} X-ray observations of NGC\,3242, 
a multiple-shell PN consisting of an inner shell with a bright, round 
rim, and an outer envelope.  
The observations have detected diffuse, soft X-ray emission confined 
within the innermost shell of NGC\,3242.  
The relatively low temperature of the hot gas, $T_X$=2.35$\times$10$^6$~K, 
compared to the expected adiabatic post-shock temperature of a stellar 
wind with a velocity of 2,400 km~s$^{-1}$ \citep{PD04} suggests that heat 
conduction has taken place.
Indeed, models including heat conduction provide a reasonable description 
of the X-ray temperature and luminosity of NGC\,3242 \citep{SSW08}.
However, the chemical abundances of the X-ray-emitting plasma are 
closer to the stellar values, suggesting little evaporation of cold nebular 
material into the hot bubble and thus contradicting the expectation of 
heat conduction.

We have compared the physical properties ($N_e$, $T_e$, $P_{\rm th}$) of 
the gas in the hot bubble with those of the gas at the inner shell rim.  
The inner shell can be described as a thin shell with a 
constant density of 2,200 cm$^{-3}$ and a thickness 15\% 
its radius, while the envelope is best described by a 
shell whose density declines $\propto$~r$^{-1/2}$ to r$^{-1/3}$.  
The gas in the hot bubble has lower thermal pressure than the gas in 
the shell rim, unless the X-ray-emitting gas is mostly confined within 
a thin shell, which is not supported by the X-ray morphology 
observed.  Comparisons with simulations favor a large filling
factor for the X-ray-emitting gas.

We also note the asymmetric distribution of the X-ray emission 
within the inner shell of NGC\,3242 and found it unlikely 
to be a result of nonuniform absorption of the X-ray emission 
within the optical shell.  
X-ray observations at higher spatial resolution and signal-to-noise ratio 
(Montez et al., in preparation) are needed to provide a sharper 
view of the distribution of the X-ray-emitting plasma within NGC\,3242.

\acknowledgments

We thank the anonymous referee for comments that helped us to improve 
this paper.
N.R.\ and M.A.G.\ are partially funded by grant AYA\,2008-01934 of 
the Spanish MICINN (Ministerio de Ciencia e Innovaci\'on).  
Y.-H.C.\ and R.A.G\ acknowledge the support of NASA grant NNG04GE63G
for this research project.
We are also very grateful to Dr.\ Karen Kwitter for providing us with 
the optical long-slit spectra of NGC\,3242.  
Some of the data presented in this paper were obtained from the 
Multimission Archive at the Space Telescope Science Institute (MAST). 
STScI is operated by the Association of Universities for Research in 
Astronomy, Inc., under NASA contract NAS5-26555. 

\facility{Facilities: HST(WFPC), XMM, KPNO(2.1m telescope).}

\clearpage

\begin{deluxetable}{lcrl}
\tablewidth{0pt}
\tabletypesize{\scriptsize}
\tablecaption{HST WFPC2 Observations of NGC\,3242}
\tablehead{
\multicolumn{1}{c}{Emission Lines} & 
\multicolumn{1}{c}{Number of images} & 
\multicolumn{1}{c}{$t_{\rm exp}$}  &
\multicolumn{1}{c}{Program ID}  \\
\multicolumn{1}{c}{}               & 
\multicolumn{1}{c}{}               & 
\multicolumn{1}{c}{[s]}       & 
\multicolumn{1}{c}{}     }
\startdata
~~~H$\alpha$     & 1 & 100 & ~~6117        \\ 
~~~He~{\sc ii}   & 1 & 160 & ~~6117       \\
~~~[N~{\sc ii}]  & 3 & 400 & ~~6117 \\
                 & 2 & 300 & ~~7501, 8773 \\
                 & 4 & 260 & ~~7501, 8773 \\
~~~[O~{\sc iii}] & 1 & 260 & ~~7501 \\
                 & 5 & 200 & ~~7501, 8773  
\enddata
\label{tab.hst}
\end{deluxetable}

%%%%%%%%%%%%%%%%%%%%%%%%%%%%%%%%%%%%%%%%%%%%%%%%%%%%%%%%%%%%%%%%%%%%%%%%%

\begin{deluxetable}{lcrrrrr}
\tabletypesize{\scriptsize}
\tablecaption{Line Strengths for NGC\,3242}
\tablewidth{0pt}
\tablehead{
\colhead{} & 
\colhead{} & 
\colhead{} &
\multicolumn{2}{c}{\underline{~~~~~~~~Inner Shell~~~~~~~~}} &
\multicolumn{2}{c}{\underline{~~~~~~~~Outer Shell~~~~~~~~}} 
\\
\multicolumn{1}{l}{Line ID}          &
\multicolumn{1}{c}{Wavelength}           &
\multicolumn{1}{c}{$f_\lambda$}        &
\multicolumn{1}{c}{~~$F$}        &
\multicolumn{1}{c}{~~$I$}        &
\multicolumn{1}{c}{~~$F$}        &
\multicolumn{1}{c}{~~$I$}        \\
\colhead{}               & 
\colhead{(\AA)}       & 
\colhead{}            & 
\colhead{}            &
\colhead{}               & 
\colhead{}       & 
\colhead{}             }
\startdata
~[O~{\sc ii}] & ~~~~3726~~~~&~~~ 0.26~~~&~~~10.8~~~&~~~11.7~~~&~~~3.7~~~&~~~4.0~~~ \\
~[O~{\sc ii}] & ~~~~3729~~~~&~~~ 0.26~~~&~~~9.7~~~&~~~10.5~~~&~~~0.3~~~&~~~0.3~~~ \\
~[S~{\sc ii}] & ~~~~4068~~~~&~~~0.20~~~&~~~0.5~~~&~~~0.5~~~& ~~~$\dots$~~~ & ~~~$\dots$~~~ \\
~[S~{\sc ii}] & ~~~~4076~~~~&~~~0.20~~~&~~~0.5~~~&~~~0.5~~~& ~~~$\dots$~~~ & ~~~$\dots$~~~ \\
~H$\delta$ & ~~~~4101~~~~&~~~ 0.18~~~&~~~26.4~~~&~~~25.6~~~&~~~27.2~~~&~~~25.5~~~ \\
~H$\gamma$ & ~~~~4340~~~~&~~0.135~~  &~~~45.5~~~&~~~47.0~~~&~~~46.2~~~&~~~48.2~~~ \\
~[O~{\sc iii}] & ~~~~4363~~~~&~~~ 0.13~~~&~~~10.9~~~&~~~11.3~~~&~~~10.1~~~&~~~10.5~~~ \\
~[Ar~{\sc iv}] & ~~~~4711~~~~&~~~0.04~~~&~~~4.8~~~&~~~4.9~~~&~~~4.0~~~&~~~4.0~~~ \\
~[Ar~{\sc iv}] & ~~~~4740~~~~&~~~0.04~~~&~~~4.0~~~&~~~4.0~~~&~~~2.9~~~&~~~2.9~~~ \\
~H$\beta$ & ~~~~4861~~~~&~~~ 0.00~~~&~~~100.0~~~&~~~100.0~~~&~~~100.0~~~&~~~100.0~~~ \\
~[O~{\sc iii}] & ~~~~4959~~~~&~~~--0.02~~~&~~~339.1~~~&~~~336.8~~~&~~~362.2~~~&~~~359.7~~~ \\
~[O~{\sc iii}] & ~~~~5007~~~~&~~~--0.03~~~&~~~1018.1~~~&~~~1007.8~~~&~~~1523.3~~~&~~~1507.9~~~ \\
~[Cl~{\sc iii}] & ~~~~5517~~~~&~~~--0.15~~~&~~~0.2~~~&~~~0.2~~~& ~~~0.3~~~ & ~~~0.3~~~ \\
~[N~{\sc ii}] & ~~~~5755~~~~&~~~--0.21~~~&~~~0.07~~~&~~~0.07~~~&~~~$\dots$~~~ & ~~~$\dots$~~~  \\
~[S~{\sc iii}] & ~~~~6312~~~~&~~~--0.30~~~&~~~0.7~~~&~~~0.6~~~&~~~0.4~~~&~~~0.4~~~  \\
~[N~{\sc ii}] & ~~~~6548~~~~&~~~--0.34~~~&~~~0.7~~~&~~~0.6~~~& ~~~$\dots$~~~ & ~~~$\dots$~~~ \\
~H$\alpha$ & ~~~~6563~~~~&~~~--0.34~~~&~~~289~~~&~~~287~~~&~~~296~~~&~~~286~~~ \\
~[N~{\sc ii}] & ~~~~6583~~~~&~~~--0.34~~~&~~~2.1~~~&~~~1.9~~~&~~~14.6~~~&~~~13.4~~~ \\
~[S~{\sc ii}] & ~~~~6716~~~~&~~~--0.36~~~&~~~0.3~~~&~~~0.3~~~&~~~1.1~~~&~~~1.0~~~ \\
~[S~{\sc ii}] & ~~~~6731~~~~&~~~--0.36~~~&~~~0.4~~~&~~~0.4~~~&~~~1.0~~~&~~~0.9~~~ \\
~[O~{\sc ii}] & ~~~~7320~~~~&~~~--0.43~~~&~~~0.5~~~&~~~0.4~~~&~~~0.05~~~&~~~0.04~~~ \\
~[O~{\sc ii}] & ~~~~7330~~~~&~~~--0.43~~~&~~~0.5~~~&~~~0.4~~~&~~~0.05~~~&~~~0.04~~~ \\
~[S~{\sc iii}] & ~~~~9069~~~~&~~~--0.64~~~&~~~3.7~~~&~~~3.1~~~&~~~22.6~~~&~~~19.2~~~ \\
~[S~{\sc iii}] & ~~~~9532~~~~&~~~--0.65~~~&~~~18.0~~~&~~~15.2~~~&~~~18.1~~~&~~~15.2~~~ \\
F(H$\beta$)
    (ergs~cm$^{-2}$~s$^{-1}$) &   &  & 3.1$\times$10$^{-11}$ &  &
2.2$\times$10$^{-12}$ &  \\
\enddata
\label{tab.flux}
\end{deluxetable}

%%%%%%%%%%%%%%%%%%%%%%%%%%%%%%%%%%%%%%%%%%%%%%%%%%%%%%%%%%%%%%%%%%%%%%%%%

\begin{deluxetable}{llrrrrr}
\tabletypesize{\scriptsize}
\tablecaption{Physical Conditions in NGC\,3242}
\tablewidth{0pt}
\tablehead{
\colhead{}  & \colhead{} & \colhead{} & 
\multicolumn{2}{c}{\underline{~~~Observed Ratios~~~}} &
\multicolumn{2}{c}{\underline{~~~~~~~~~Value~~~~~~~~~}} \\
\colhead{Physical Parameter}          &
\colhead{Ion}        &
\colhead{Line Ratios}           &
\colhead{Inner Shell}        &
\colhead{Outer Shell}        &
\colhead{Inner Shell}        &
\colhead{Outer Shell}  }
\startdata
$T_{e}$ & [N~{\sc ii}]    &  (6548+6583/5755         & 41.6~~~~~ & $\dots$~~~~~~~ & 14,700 K~~ & $\dots$~~~~~~  \\
$T_{e}$ & [S~{\sc iii}]   &  (9069+9532)/6312        & 32.0~~~~~ & 106.6~~~~~ & 14,040 K~~ & 8,070 K~~ \\
% $T_{e}$ & [S~{\sc ii}] &  (6716+6731)/(4068+4076) & 0.61 & $\dots$ & $\dots$ & $\dots$ \\
$T_{e}$ & [O~{\sc iii}]   &  (4959+5007)/4363 & 124.9~~~~~ & 189.0~~~~~ & 11,880 K~~ & 10,400 K~~ \\
$T_{e}$ & [O~{\sc ii}]    &  (3726+3729)/(7320+7330) & 19.5~~~~~ & 43.3~~~~~ & 10,100 K~~ & 9,860 K~~ \\ 
\hline
$N_{e}$ & [Ar~{\sc iv}]   &  4711/4740               & 1.18~~~~ & 1.37~~~~ & 2,250 cm$^{-3}$ & 400 cm$^{-3}$ \\
$N_{e}$ & [S~{\sc ii}]    &  6716/6731               & 0.71~~~~ & 1.13~~~~ & 2,200 cm$^{-3}$ & 340 cm$^{-3}$ \\

\enddata
\label{tab.tn}
\end{deluxetable}

\clearpage

\end{document}